\documentclass[amstex,aps,showpacs,prl,twocolumn]{revtex4}%
\usepackage{graphicx}
\usepackage{amsmath}
\usepackage{bm}%
\usepackage{amsfonts}%
\usepackage{amssymb}

\begin{document}

\title{Observation of diffractive orbits in the spectrum of excited NO in a magnetic field}

\author {A.\ Matzkin}
\affiliation{Laboratoire de Spectrom\'{e}trie physique (CNRS
Unit\'{e} 5588), Universit\'{e} Joseph-Fourier Grenoble-I, BP 87,
38402 Saint-Martin, France}
\author{M. Raoult}
\affiliation{Laboratoire Aim\'{e} Cotton (CNRS Unit\'{e} 3321),
Universit\'{e} de Paris-Sud, 91405 Orsay, France}
\author {D. Gauyacq}
\affiliation{Laboratoire de Photophysique mol\'{e}culaire (CNRS
Unit\'{e} 3361), Universit\'{e} de Paris-Sud, 91405 Orsay, France}

\begin{abstract}
We investigate the experimental spectrum of excited NO molecules
in the diamagnetic regime and develop a quantitative semiclassical
framework to account for the results. We show the dynamics can be
interpreted in terms of classical orbits provided that in addition
to the geometric orbits, diffractive effects are appropriately
taken into account. We also show how individual orbits can be
extracted from the experimental signal and use this procedure to
reveal the first experimental manifestation of inelastic
diffractive orbits.
\end{abstract}

\maketitle

%PACS 32.60.+i 03.65.Sq 32.55.Be 05.45.Mt
%33.55.Be stark and zeeman for MOLECULES
%32.60.+i stark and zeeman for ATOMS

Periodic orbit (PO) theory \cite{gutzwiller90} has been a very
successful quantitative and qualitative tool in the analysis of
systems displaying strong quantum fluctuations. In particular, the
complex photoabsorption spectrum of excited hydrogen atoms in a
static magnetic field was interpreted within closed orbit theory
(COT) in terms of periodic orbits \emph{closed} at the nucleus
\cite{du delos88}. However standard PO theory was shown to be
insufficient for systems containing a scatterer smaller or
comparable to one de Broglie wavelength, because the PO's may hit
the scatterer, thereby producing \emph{diffractive} orbits
\cite{vattay}. In certain mesoscopic devices diffractive orbits
crucially influence the spectrum \cite{heller99}, whereas in
non-hydrogenic Rydberg atoms in static fields the atomic core acts
as a scatterer producing combination orbits \cite{dando etal95}.\
This was experimentally observed in the photoabsorption spectrum
of helium, which departs slightly from the hydrogenic case
\cite{karremans etal 1998}. In the case of Rydberg molecules, core
effects are expected to be more important: the reason is that the
diffractive scatterer has an internal structure (the quantum
states of the molecular core).\ Indeed, model calculations
recently predicted that Rydberg molecules in external fields would
display a novel type of diffractive orbit produced by inelastic
scattering of the PO's on the core \cite{matzkin etal2002}.

\begin{figure}[b]
\includegraphics[height=1.7in,width=2.1in]{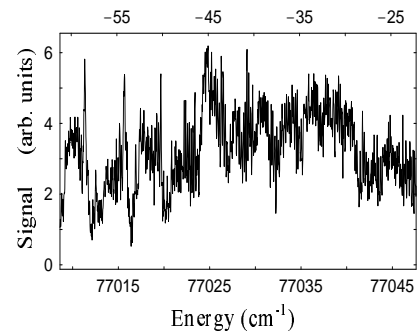}
\caption[]{Experimental spectrum for the transition from the
$A^{2}\Sigma ^{+},$ $N=3$, $M_{N}=-3,...,3$
 states of NO in a static magnetic field of strength $B=0.93$ T.
 The lower scale gives the energy relative to the ground state
of NO. The origin of the upper scale corresponds to the field-free
ionization threshold for levels converging to the $N^{+}=1$
rotational state of NO$^{+}$.\label{fig1}}
\end{figure}

In this work, we address this problem by investigating the
experimental spectrum of excited NO molecules in a magnetic
field.\ We show the apparently complex spectrum can be well
understood in terms of closed orbits provided molecular core
effects are appropriately taken into account; in particular, we
shall see COT calculations correctly reproduce the Fourier
transform of the experimental spectrum, in a regime in which
quantum calculations have yet to be undertaken. We will also
demonstrate how individual orbits may be extracted from the
experimental results by using a modified Fourier tranform, and use
this technique to show the first experimental manifestation of an
\emph{inelastic diffractive} orbit in a physical system.

Most previous studies of atoms in a magnetic field have relied on
scaled-energy spectroscopy, in which the field $B$ is varied with
the Rydberg
electron energy $E$ so as to keep the scaled energy $\epsilon=E(B/B_{0}%
)^{-2/3}$ constant, with $B_{0}=2.35\times10^{5}$ T. Then the
classical motion is unchanged at each point of the spectrum and
each orbit contributes as a peak in the Fourier transformed (FT)
spectrum. However, such a scaling law does not hold, albeit
approximately, for molecules.\ We have therefore observed the
Rydberg states of NO at a constant magnetic field of 0.93 T,
varying the excitation energy of the probe laser.\ The experiment,
carried out in a ''magnetic bottle'' time of flight spectrometer,
was described elsewhere \cite{gauyacq}, but in short: a 2-photon
highly selective transition with a pump laser brings the molecule
from the ground state $X^{2}\Pi_{3/2}$ (in the vibrational level
$v"=0$) into the single rotational Zeeman sublevel $N=3$
$M_{S}=1/2$ of the first electronic excited state
$A^{2}\Sigma^{+},$ $v^{\prime}=1$; $N$ is the total angular
momentum exclusive of spin and the constant spin number $M_{S}$
will henceforth be disregarded. The sublevels $M_{N}=-3$ to $3$
(where $M_{N}$ is the projection of $N$ on the field axis) were
not resolved but their relative population is known. In a second
step, the probe laser, linearly polarized parallel to the field
axis, excites the molecules to high-lying Rydberg states having
also $v=1$ (so we can neglect vibrational couplings).

These excited states are further ionized via vibrational
autoionization into the ground state of the NO$^{+}$ core. The
resulting spectrum,\thinspace which lies higher in energy than
those previously published in \cite{gauyacq} (in which
nonperturbative diamagnetic effects could be neglected), is shown
in Fig. 1. The FT of the experimental spectrum is shown in Fig
2(a), whereas Fig.\ 2(b) displays the semiclassical calculations
which will be detailed below. Unlike a spectrum taken at constant
$\epsilon,$ we do not expect in the present case to correlate each
peak with a single orbit, because the period of an orbit changes
with the energy.\ Before turning to the extraction of individual
orbits, we detail the mechanism giving rise to the peaks in the FT
spectrum.

\begin{figure}[t]
\includegraphics[height=2in,width=2.2in]{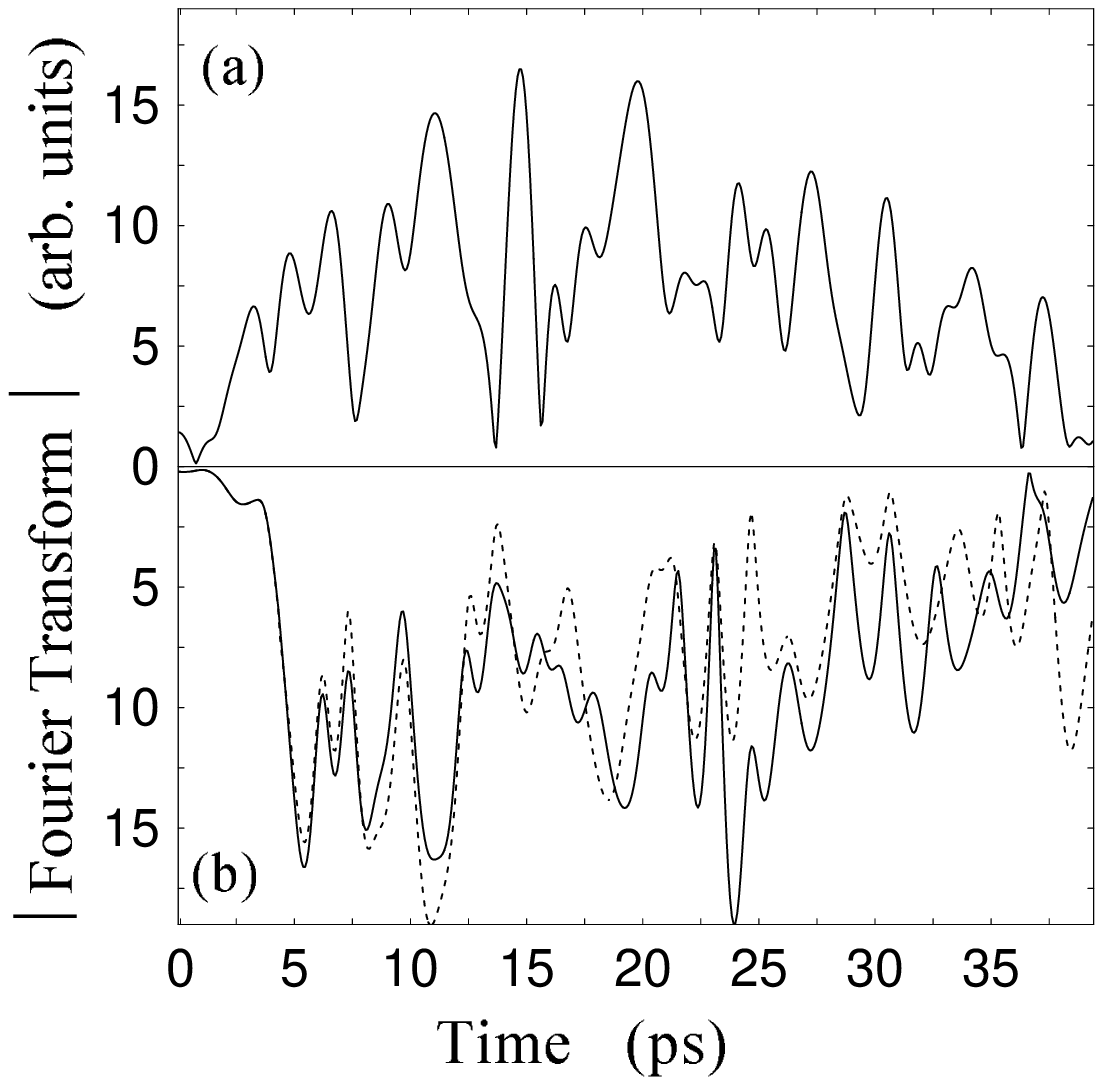}
\caption[]{(a) Fourier-transform of the experimental spectrum. The
first peak (marked with a star) is a residual oscillation from the
background due to the redistribution of the oscillator strength
into molecular Zeeman structures at low energies \cite{raoult
etal2003}. (b) Closed-orbit semiclassical calculations with (solid
line) and without (dotted line) the diffractive contributions. \
An independent calculation was performed for each value of
$M_{N}$; each result was then multiplied by the relative weight of
the $M_{N}$ component before taking the Fourier transform. Note
that the wide peaks are not correlated with single closed orbits
but arise from a coherent superposition of orbits at different
energies. The shortest contributing orbit is an orbit
perpendicular to the field with a period of 4.22 ps.\label{fig2}}
\end{figure}

\begin{figure}[b]
\includegraphics[height=2.75in,width=3.in]{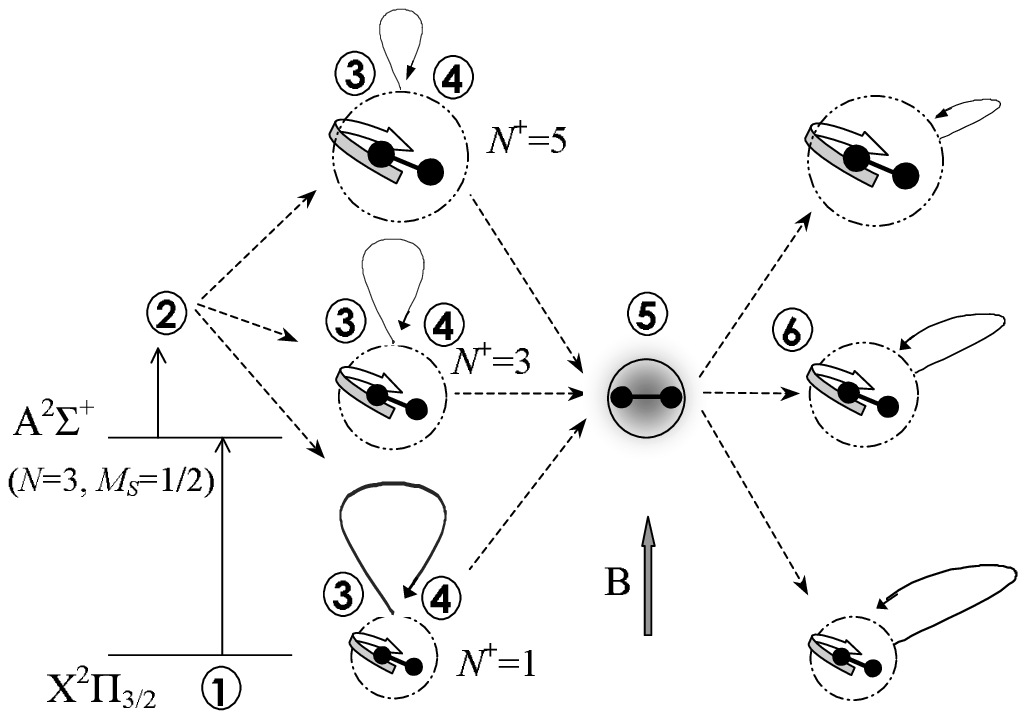}
\caption[]{Schematic diagram of the excitation dynamics of the
Rydberg states of NO in a magnetic field. (1) A pump laser excites
the molecules from the ground state $X^{2}\Pi _{3/2}$ to the
intermediate state $A^{2}\Sigma ^{+},$ $N=3,$ via a 2-photon
transition. A high-resolution probe laser then excites the
molecules from the intermediate state to Rydberg levels with
orbital momentum $l=1$ and $l=3$. (2) Following this excitation
the Rydberg electron uncouples from the molecular
 core which is left freely rotating with rotation quantum numbers $N^{+}%
=1,$ 3 or $5$. (3) As the Rydberg electron leaves the core region
its dynamics
 is described using classical trajectories; the molecular energy is partitioned
  between the Rydberg electron and the core: as the core rotation increases
(pictured by a larger core), the energy of the Rydberg electron
decreases (symbolized by smaller orbits).\ The core region is less
than 5 a.u. large, whereas even the shortest orbit associated with
a highly rotating core in state $N^{+}=5$ extends beyond 1500 a.u.
(4) Some trajectories return to the core, which following the
terminology of diffractive PO theory we call the \emph{geometric
closed orbits}; they are the same orbits that exist in the
diamagnetic hydrogen atom (i.e. in the absence of the core). (5)
As the Rydberg electron enters the core region, its dynamics
recouples with the core.\ The waves carried by the geometric CO's
overlap with the initially-dipole excited waves, producing
fluctuations in the oscillator strength. (6) The electron scatters
off the core, uncouples, and follows again classical trajectories;
during the scattering process, the electron may have exchanged
energy with the core.\ Eventually, some trajectories return once
more to the core; these \emph{diffractive orbits}, composed of 2
geometric CO's connected by elastic or inelastic core-scattering,
produce further fluctuations in the oscillator strength.}
\end{figure}

\begin{table}[tbp] \centering
\begin{tabular}
[c]{|c|c|c|c|}\hline $N^{+}$ & 1 & 3 & 5\\\hline $\epsilon$ &
$-1.1\longrightarrow-0.4$ & $-1.5\longrightarrow-0.7$ &
$-2.1\longrightarrow-1.4$\\
$\left|  \mathcal{Q}_{N^{+}}^{G}\right|  $ & $0.133$ & $0.304$ & $0.225$\\
$\left|  \mathcal{Q}_{N^{+},3}^{D}\right|  $ & $0.004$ & $0.264$ &
$0.003$\\\hline
\end{tabular}
\caption{For each state of the core, we give the range of the
scaled energy $\epsilon $
 of the Rydberg electron as the probe laser is scanned. When $N^{+}%
=5$, the very low values of $\epsilon $ correspond to the near
integrable
 regime. For $N^{+}=3,$ the dynamics of the Rydberg electron is mostly
regular.\ For $N^{+}=1$, the electron enters the chaotic regime:
at $\epsilon =-0.4,$ about 50 \% of the classical phase space is
chaotic \cite{friedrich wintgen89}. We also give the modulus of
$\mathcal{Q}^{G}$ and $\mathcal{Q}^{D}$ for $N^{+\prime}=3$
 and for $M_{N}=-1;$ these values give a rough
 idea of the PO-independent part of
the fluctuations strength associated with the different geometric
and diffractive contributions.}
\end{table}

\begin{figure}[b]
\includegraphics[height=3.35in,width=2.5in]{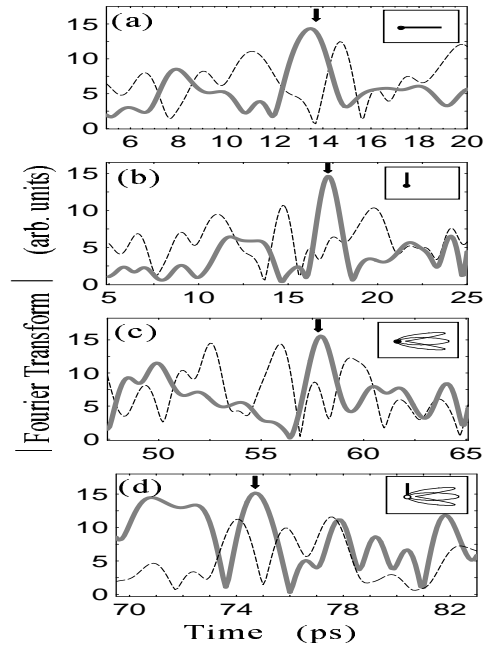}
\caption[]{Extraction of individual orbits from the experimental
spectrum using the chirped FT procedure. In the present
illustration, we show the results for 4 orbits which exist at an
excitation energy $E_{0}=-26.91$ cm$^{-1}$ below the field free
$N^{+}=1$ ionization threshold. (a) Extraction of the second
return of the perpendicular orbit ($R_{2}$ in the nomenclature
given e.g. in \cite{main99}), associated with the core in state
$N^{+}=5$; the corresponding scaled energy of this orbit is
$\epsilon =-1.49$. \emph{Broken curve}: standard FT of the
experimental spectrum, shown in Fig.\ 1. \emph{Thick curve}:
chirped FT of the experimental spectrum obtained by calculating
the function $\xi (t;E)$ appearing in Eq. (\ref{4}) for this
orbit.\ The arrow indicates the calculated classical period of
$R_{2},$ which agrees very well with the position of the peak
appearing in the CFT. (b) CFT extraction of the orbit $V_{1}$,
parallel to the field axis, associated with the core in state
$N^{+}=3,$ at $\epsilon =-0.85$. (c) Extraction of the $R_{3}^{1}$
orbit associated with the core in state $N^{+}=1$; orbits
associated with $N^{+}=1$ are generically more difficult to
extract from the experimental signal because the oscillator
strength tends to be dominated by orbits associated with $N^{+}=5$
and $3$ (see Table I). However, at this energy, corresponding to
$\epsilon =-0.48,$ the classical amplitude $A_{R_{3}^{1}}$ is very
strong (because $\epsilon $ is close to the bifurcation energy at
which this orbit is created \cite{mao delos92}), thereby allowing
the signal to pick up its contribution. (d) By applying the CFT
procedure to the action expansion
$S_{V_{1}}(E_{N^{+}=3})+S_{R_{3}^{1}}(E_{N^{+}=1}),$ we can also
extract the \emph{diffractive orbit }formed by the combination of
the orbits extracted in (b) and (c); the period of the diffractive
orbit is given by summing the periods of the two geometric CO's.
The shape of the orbits is given in the upper right-hand side of
each plot ($B$ is in the vertical direction). \label{fig4}}
\end{figure}

Since in the $A^{2}\Sigma^{+}$ state the orbital momentum of the
Rydberg electron is a mixture of both $l=0$ and $l=2$, the probe
laser excites to final states having $l=1,3$ and $N=2,3,4$.
Photo-excitation takes place in the molecular frame (in which the
Rydberg electron is coupled to the molecular core), but following
excitation the electron uncouples from the core and $N$ is no
longer conserved, although its projection $M_{N}$ is
\cite{monteiro taylor90}. The molecule is then described as an
electron orbiting around the molecular core NO$^{+}$ in
well-defined rotational states $N^{+}=1,3$ or $5.$ For each core
configuration, the electron's wavefunction is propagated
semiclassically along classical trajectories (see Fig. 3). The
energy of the Rydberg electron is obtained from the energy
partition%

\begin{equation}
E_{N^{+}}=E_{N^{+}=1}+B_{r}\left[  N^{+}(N^{+}+1)-2\right]  +\frac{m_{l}}%
{2}B/B_{0},\label{1}%
\end{equation}
where $B_{r}$ is the rotational constant of NO$^{+}$. Disregarding
the linear Zeeman shift, we have three main dynamical regimes for
the Rydberg electron, each associated with a different rotational
core state.\ Table I gives, for each value of $N^{+}$ the scaled
energy ranges corresponding to the experimental spectrum.

At each energy $E_{N^{+}},$ some of the trajectories are turned
back by the field and return to the core; those trajectories are
the \emph{geometric closed orbits}. Their contribution to the
oscillator strength $f(E)$ is given by
\cite{raoult etal2003}%
\begin{equation}
\sum_{N^{+}}\mathcal{Q}_{N^{+}}^{G}\sum_{k}\mathcal{A}_{k}(E_{N^{+}}%
)\exp\left[  i\mathcal{S}_{k}(E_{N^{+}})\right]  . \label{2}%
\end{equation}
$\mathcal{A}_{k}$ and $\mathcal{S}_{k}$ are 3-dimensional
quantities respectively linked to the 2-dimensional classical
amplitude $A_{k}$ and action $S_{k}$ of the $k$th closed orbit
\cite{GD92} ($\mathcal{S}_{k}$ includes the Maslov index
$\mu_{k}(E_{N^{+}})$); those quantities depend on the energy of
the Rydberg electron, i.e. on the core state and on the excitation
energy. $\mathcal{Q}_{N^{+}}^{G}$ is a quantity depending on the
dipole transition factor from the $A^{2}\Sigma^{+}$ to the excited
states, and on the transformation elements from the coupled to the
uncoupled frame; some values are given in Table I.

When the semiclassical wavefunction carried by the geometric
closed orbits returns to the core, it diffracts on the scatterer
(Fig. 3).\ This diffraction process, produces newly outgoing
trajectories $q,$ which eventually return to the core;
the resulting diffractive contribution to $f(E)$ is given by%
\begin{align}
&  \sum_{N^{+},N^{+\prime}}\mathcal{Q}_{N^{+},N^{+\prime}}^{D}\sum
_{k,q}\mathcal{A}_{k}(E_{N^{+}})\mathcal{A}_{q}(E_{N^{+\prime}})\nonumber\\
&  \times\exp i\left[  \mathcal{S}_{k}(E_{N^{+}})+\mathcal{S}_{q}%
(E_{N^{+\prime}})\right]  . \label{3}%
\end{align}
$\mathcal{Q}_{N^{+},N^{+\prime}}^{D}$ acts as a global diffraction
coefficient; unlike diffraction in quantum billiards
\cite{vattay,heller99}, $\mathcal{Q}_{N^{+},N^{+\prime}}^{D}$ does
not depend here on the local geometry of the scatterer but solely
on the quantum properties of the core (i.e. on the core-induced
phase-shifts, known as molecular quantum defects; the value of the
6 independent quantum defects for NO is given in \cite{gauyacq}).\
The matching between the semiclassical waves and the core
wavefunction takes place in the stationary phase approximation.
The oscillator strength contains both elastic
($N^{+}=N^{+\prime})$ and inelastic ($N^{+}\neq N^{+\prime})$
diffractive terms. In the one core-scatter approximation, the
total contribution to the fluctuations in the oscillator strength
is obtained by summing the imaginary parts of Eqs. (\ref{2}) and
(\ref{3}).

The experimental or semiclassically calculated FT spectra $\left|
\mathcal{F}(t;\xi)\right|  $ are respectively obtained by Fourier
transforming the photoelectron signal or the oscillator
strength for which we shall use the same notation $f(E)$, hence%
\begin{equation}
\mathcal{F}(t;\xi)=\int_{E_{\min}}^{E_{\max}}f(E)\exp\left(  -i\xi
(t,E)\right)  dE \label{4}%
\end{equation}
where $\xi(t,E)=Et,$ and we set the energy variable relative to
the $N^{+}=1$ ionization threshold as $E=E_{N^{+}=1}$. The
semiclassical calculations were undertaken by applying Eqs.
(\ref{2}) and (\ref{3}). Since at each point $E_{0}$ of the
spectrum the orbits and therefore their amplitude and action
change, we have determined the classical quantities $A_{k},$
$S_{k}$, $\mu_{k}$ on an energy grid for the primitive closed
orbits and their repetitions, keeping track of bifurcations as the
energy increases. We then interpolated those functions so as to
obtain classical quantities varying with the energy. This
procedure, which is tractable for short orbits provided the scaled
energy is not too high, yields a good agreement with the
experimental FT\ spectrum, as seen on Fig.\ 2(b). The
semiclassical calculations allow to single out contributions from
the ensemble of geometric or diffractive orbits associated with
given core states. Full details will be given elsewhere
\cite{raoult etal2003}.

Eq. (\ref{4}) is also at the basis of the extraction of
\emph{individual orbits} from the experimental spectrum provided
$\xi(t,E)$ is appropriately modified so as to counterbalance the
varying values of the classical quantites.\ Indeed in the
neighborhood of some energy $E_{0}$ a classical orbit $k$
contributes locally as a sinusoidal oscillation of period
$T_{k}(E_{0})$. But as the energy is varied, the classical motion
changes, turning the sinusoidal oscillation into a chirped
modulation. It has been shown very recently \cite{delos cft2002}
that a chirped Fourier Transform (CFT) converts a chirp modulation
into a sine oscillation, thus giving a sharp peak in the FT
spectrum. A CFT is defined by setting
$\xi(t,E)=t(E-E_{0})+b(E-E_{0})^{2}+c(E-E_{0})^{3}+i\ln\alpha(E-E_{0})$
in Eq. (\ref{4}). $b$ and $c$ are numerical coefficients and
$\alpha$ is a function, which depend on the orbit to be extracted:
to extract an orbit $k$ at some energy $E_{0},$ we Taylor expand
the action $S_{k}(E)$.\ To first order, we have
$\partial_{E}S(E_{0})=T_{k}(E_{0}),$ i.e.\ the period of the
orbit. Therefore by setting $b=-\partial_{E}^{2}S_{k}(E_{0})/2$
and $c=-\partial _{E}^{3}S_{k}(E_{0})/6$ in the CFT exponent
$\xi,$ we can expect to effectively linearize the signal around
$E_{0}$ so that the CFT yields a peak at $t=T_{k}(E_{0})$. The
peak can be enhanced by using a function $\alpha$ to compensate
for the signal amplitude variation; we have used the inverse of
our interpolated functions $A_{k}(E)$. As discussed in great
details in \cite{delos cft2002}, spurious peaks inevitably appear
in the CFT spectrum.\ We proceed in the following manner: using
purely classical calculations, we compute on the one hand the CFT
of the experimental signal with the relevant parameters $a,b,$ and
$\alpha,$ and on the other hand we determine the period
$T_{k}(E_{0})$; the consistency of the procedure is ensured by the
appearance of a peak at the computed value $t=T_{k}(E_{0})$ when
the orbit is succesfully extracted from the signal. An
illustration of the extraction procedure is given in Fig. 4. For a
fixed energy $E_{0}$ we extract one CO associated with each core
state (panels (a)-(c)). Panel (d) shows the extraction of the
diffractive CO formed by the combination of the orbits extracted
in (b) and (c).

In summary, we have investigated the experimental spectrum of NO
in the diamagnetic regime and interpreted the results in terms of
geometric and diffractive classical orbits, and we have shown how
individual orbits can be extracted from the experimental signal.


\begin{thebibliography}{9}

\bibitem {gutzwiller90}M. C.\ Gutzwiller, \emph{Chaos in classical and quantum
mechanics} (Springer, New-York, 1990).

\bibitem {du delos88}M.\ L.\ Du and J.\ B.\ Delos, Phys. Rev. A \textbf{38},
1913 (1988).

\bibitem {vattay}G.\ Vattay, A.\ Wirzba and P.E\
Rosenqvist, Phys.\ Rev.\ Lett. \textbf{73}, 2304 (1994).

\bibitem {heller99}J. S. Hersch, M. R. Haggerty, and E. J. Heller,
Phys.\ Rev.\ Lett. \textbf{83}, 5342 (1999).

\bibitem {dando etal95}P.\ A.\ Dando, T. S.\ Monteiro, D.\ Delande and
K.\ T.\ Taylor, Phys. Rev. Lett. \textbf{74}, 1099 (1995).

\bibitem{karremans etal 1998}K. Karremans, W. Vassen, and W. Hogervorst,
 Phys. Rev. Lett. \textbf{81}, 4843 (1998).

 \bibitem {matzkin etal2002}A. Matzkin, P. A. Dando, and T. S. Monteiro,
 Phys. Rev. A \textbf{66}, 013410 (2002).

\bibitem {gauyacq}D.\ Gauyacq, M. Raoult and N.\ Shafizadeh, in C.
Sandorfy (Ed.), \emph{The Role of Rydberg States in Spectroscopy
and Photochemistry} (Kluwer, Dordrecht, 2002); M. Raoult, S.
Guizard and D. Gauyacq, J. Chem. Phys. \textbf{94}, 7046 (1991).

\bibitem {raoult etal2003}M. Raoult, D. Gauyacq, S. Guizard and A.
Matzkin, in preparation.

\bibitem {monteiro taylor90}T.\ S.\ Monteiro and K.\ T.\ Taylor, J. Phys. B
\textbf{23}, 427 (1990).

\bibitem {GD92}J.\ Gao, J.B.\ Delos, and
M.\ Baruch, Phys.\ Rev.\ A \textbf{46}, 1449 (1992).

\bibitem {friedrich wintgen89}H.\ Friedrich and D.\ Wintgen,
Phys.\ Rep.\ \textbf{183}, 37 (1989).

\bibitem {delos cft2002}S. Freund, R. Ubert, E. Fl\"{o}thmann,
K. Welge, D. M. Wang and J. B. Delos Phys. Rev. A \textbf{65},
053408 (2002).

\bibitem {main99}J. Main, Phys. Rep. \textbf{316},
233 (1999).

\bibitem {mao delos92}J.–M. Mao and J. B. Delos Phys. Rev. A \textbf{45}, 1746
(1992).


\end{thebibliography}
\end{document}